\documentclass[11pt]{article} 

\usepackage[ansinew]{inputenc}
\usepackage{amsmath, amssymb, graphics, amsthm}
\usepackage{epsfig}
\usepackage{color}
\usepackage{fancyhdr} 

\usepackage[normalem]{ulem}

\makeatletter
\@addtoreset{equation}{section}
\makeatother

\newcommand{\be}{\begin{equation}}
\newcommand{\ee}{\end{equation}}
\newcommand{\ba}{\begin{eqnarray}}
\newcommand{\ea}{\end{eqnarray}}

\def\pb#1{\rlap{\lower1.5ex\hbox{$\longleftarrow$}}{#1}}
\def\dpb#1{\rlap{\lower1.5ex\hbox{$\Longleftarrow$}}{#1}}
\def\spb#1{\rlap{\lower1.0ex\hbox{$\leftarrow$}}{#1}}
\def\sdpb#1{\rlap{\lower1.0ex\hbox{$\Leftarrow$}}{#1}}

\DeclareMathOperator{\sign}{sign}
\DeclareMathOperator{\acosh}{acosh}
\newcommand{\del}{\partial}
\renewcommand{\Re}{\operatorname{Re}}
\renewcommand{\Im}{\operatorname{Im}}

\title{{\sf Imaginary action, spinfoam asymptotics and the ``transplanckian'' regime of loop quantum gravity}} 
\author{
{\sf N. Bodendorfer}$^{1}$\thanks{{\sf 
norbert@gravity.psu.edu}},
{\sf Y. Neiman}$^{1}$\thanks{{\sf 
yashula@gmail.com}}
\\
{\sf $^1$ Institute for
Gravitation and the Cosmos \& Physics
  Department,}\\
{\sf   Penn State, University Park, PA 16802, U.S.A.}\\
}
\date{{\small\sf \today}}

\begin{document} 

\maketitle

{\sf

\begin{abstract}
It was recently noted that the on-shell Einstein-Hilbert action with York-Gibbons-Hawking boundary term has an imaginary part, proportional to the area of the codimension-2 surfaces on which the boundary normal becomes null. We discuss the extension of this result to first-order formulations of gravity. As a side effect, we settle the issue of the Holst modification vs. the Nieh-Yan density by demanding a variational principle with suitable boundary conditions. We then set out to find the imaginary action in the large-spin 4-simplex limit of the Lorentzian EPRL/FK spinfoam. It turns out that the spinfoam's effective action indeed has the correct imaginary part, but only if the Barbero-Immirzi parameter $\gamma$ is set to $\pm i$ {\it after} the quantum calculation. We point out an agreement between this effective action and a recent black hole state-counting calculation in the same limit. Finally, we propose that the large-spin limit of loop quantum gravity can be viewed as a high-energy ``transplanckian'' regime.\\
\mbox{}\\
\mbox{}

\end{abstract}

}

\newpage

\section{Introduction}

The verification of the correct classical limit of a theory of quantum gravity, i.e. general relativity (GR), is the most basic and commonly agreed upon requirement we demand from it. For this, it is mandatory to have a thorough understanding of the classical theory. 
An interesting feature of the classical Einstein-Hilbert action for general relativity, which can serve as a non-trivial test of the classical limit(s) of a quantum theory\footnote{A quantum field theory can have more than one classical limit. The oldest example is wave/particle duality: quantum electrodynamics can be described in one limit by a classical field, and in another limit by classical particles (photons). In section \ref{sec:Discussion}, we will encounter two candidate classical limits of loop quantum gravity.}, was recently pointed out in \cite{NeimanOnShellActions, NeimanTheImaginaryPart}. There, it was shown that the on-shell Einstein-Hilbert action with York-Gibbons-Hawking boundary term \cite{YorkRoleOfConformal, GibbonsActionIntegralsAnd} for a finite region always has an imaginary part, proportional to the area of the codimension-2 surfaces where the boundary normal becomes null. The imaginary part arises from analytically continuing the normal's angle near such surfaces, which one must do to avoid a pole singularity\footnote{\label{ftn:divergence}It is well known that the York-Gibbons-Hawking boundary term diverges for non-compact spatial slices; see \cite{AshtekarAsymptoticsAndHamiltoniansFour} for a discussion. This divergence is however conceptually different and not the cause for the imaginary part of the action.} in the York-Gibbons-Hawking boundary integral. 

An imaginary action in GR has been discussed long ago by Gibbons and Hawking \cite{GibbonsActionIntegralsAnd} in the context of a stationary black hole Wick-rotated to Euclidean spacetime. After correcting for an infinite constant and some terms related to conserved charges, the imaginary action of \cite{GibbonsActionIntegralsAnd} yields the Bekenstein-Hawking entropy. The results of \cite{NeimanOnShellActions,NeimanTheImaginaryPart} indicate that an imaginary part is a much more general feature of gravitational actions, and that we should look for it in the classical limits of candidate quantum gravity theories. The general study of actions in finite regions, in particular for gravity, is motivated in more detail in section \ref{sec:FiniteRegions}. As discussed in \cite{NeimanTheImaginaryPart, NeimanAsymptotic}, the imaginary part of the action does not contribute to the variational principle. It is nevertheless interesting, due to its close relation with gravitational entropy.

In the classical limit, the (effective) on-shell action is related to transition amplitudes through the path integral formalism. Therefore, models based on path-integral quantization are well suited for testing the above-mentioned feature of the GR action. Loop quantum gravity (LQG) \cite{RovelliQuantumGravity, ThiemannModernCanonicalQuantum} is a candidate theory of quantum gravity that comes both in a Hamiltonian formulation and in a path integral framework, known as spinfoam models. The currently most studied spinfoam models are the EPRL/FK models \cite{FreidelANewSpin, EngleLoopQuantumGravity} for which many results are known. We refer to \cite{PerezTheSpinFoam} for a recent review. In particular, one can study ``semiclassical'' coherent boundary states in the limit of large spins (corresponding to large areas). In our context, then, one would like to recover the imaginary part of the GR action from the spinfoam amplitudes for such states. We consider this issue in section \ref{sec:SpinFoam}. We find that the imaginary action is indeed recovered, but only if one sets the Barbero-Immirzi parameter $\gamma$ to $\pm i$ at the end of the calculation. This is intriguing, since $\gamma = \pm i$ corresponds to self-dual Ashtekar-Barbero variables. We stress, however, that setting $\gamma$ to $\pm i$ is a formal procedure, since there is currently no detailed understanding of the quantum theory with $\gamma$ non-real.  

On a different route, one might be worried that the boundary term, and especially its imaginary part, might depend on the precise classical formulation. 
The results of \cite{NeimanOnShellActions, NeimanTheImaginaryPart} were derived in a second-order framework. It is of interest to see if they survive in the first-order Palatini action. In particular, the current spinfoam models are based on a modification of the Palatini action, known as the Holst action \cite{HojmanParityViolationIn,HolstBarberosHamiltonianDerived}. In the literature, one can find a variety of boundary terms attributed to first-order actions. The boundary term cited for the Palatini action is sometimes equivalent to the York-Gibbons-Hawking term \cite{ObukhovThePalatiniPrinciple,WielandComplexAshtekarVariables,BianchiHorizonEnergyAs}, and sometimes not \cite{AshtekarAsymptoticsAndHamiltoniansFour,CorichiSurfaceTermsAsymptotics}. The latter is suggested in \cite{AshtekarAsymptoticsAndHamiltoniansFour} to be better behaved, since it remains finite for asymptotically flat spatial slices. As for the Holst modification, it may be given as just a curvature-proportional Lagrangian \cite{WielandComplexAshtekarVariables,BianchiHorizonEnergyAs}, or just a torsion-proportional boundary term \cite{MercuriFermionsInThe}, or a combination of the Lagrangian with a slightly different boundary term \cite{CorichiSurfaceTermsAsymptotics}. The pure-boundary-term version is the integral of the Nieh-Yan topological density, thus implying that the Holst modification is topological\footnote{We thank the paper's referees for pointing out several of these references.}.

In section \ref{sec:BoundaryTerm}, we argue for a single correct form of the first-order action. Our action has a Palatini boundary term as in \cite{WielandComplexAshtekarVariables,BianchiHorizonEnergyAs}, and a Holst part consisting of \emph{both} the Lagrangian from \cite{WielandComplexAshtekarVariables,BianchiHorizonEnergyAs} and the boundary term from \cite{MercuriFermionsInThe}. Our criteria for selecting the action are gauge invariance and a variational principle where only the boundary intrinsic metric must be kept fixed. The resulting action is equal on-shell to the second-order action, and therefore has the same imaginary part.

Section \ref{sec:Discussion} is devoted to discussion, focusing on the spinfoam results from section \ref{sec:SpinFoam}. We propose to view these results within a physical interpretation of the large-spin limit of LQG as a ``transplanckian'' high-energy regime. In this framework, we make contact between the perturbative running of $\gamma$ in effective field theory \cite{BenedettiPerturbativeQuantumGravity} and our proposal to set $\gamma = \pm i$ in the spinfoam calculation. We also place in this context some recent work \cite{FroddenBlackHoleEntropy,BSTI} on black hole entropy within loop quantum gravity. In particular, we point out a first-ever agreement within LQG between a black hole state counting and the Bekenstein-Hawking entropy as derived from an appropriate effective action. Some of our remarks are speculative and should be taken with due care. 

\section{Why finite regions and closed boundaries?}

\label{sec:FiniteRegions}

In field theory, one always works with some sort of boundary data. The action principle restricts us to boundaries that are closed hypersurfaces, enveloping some region of spacetime. It is often convenient to place the boundary hypersurface at asymptotic infinity, either in whole or in part. One example is the calculation of S-matrix elements, where the ``in'' and ``out'' states are given on the asymptotic boundary of Minkowski space. Another example is when the initial and final states are given on two constant-time slices. These slices intersect at spatial infinity, thus forming a closed boundary.

On the other hand, gravity actually forces us to consider physics in finite regions. This is because spacetime curvature together with causality can make asymptotic infinity inaccessible from certain locations. The prime example is an observer inside a black hole. At the cosmological scale, it appears that \emph{all} observers are in a similar predicament, due to the universe's accelerating expansion. In both scenarios, one encounters the notion of a finite entropy associated with the causal horizon.

\begin{figure}%
\centering%
\includegraphics[scale=0.75]{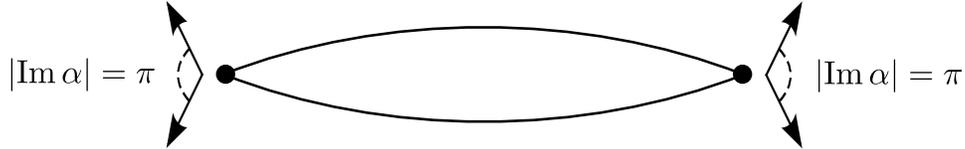} \\
\caption{A purely spacelike closed boundary, composed of two intersecting hypersurfaces. The full circles denote the corner surface. The arrows indicate the two boundary normals at each intersection point. A continuous boost between these two normals involves two signature flips. As a result, the ``corner angle'' has an imaginary part with magnitude $\pi$.}
\label{fig:spacelike_boundary} 
\end{figure}

Given these circumstances, it is interesting to learn about any peculiarities of field theory that are specific to finite regions with Lorentzian causality. An important source of such peculiarities is the presence of ``signature-flip'' surfaces. These are codimension-2 surfaces where the boundary changes its signature from spacelike to timelike (or vice versa), and momentarily becomes null. Any closed boundary must contain such surfaces. The flip surfaces may be ``hidden'' in topological corners, where the boundary ``makes a sharp turn'' in a non-differentiable manner. For example, a closed boundary composed of two spacelike hypersurfaces, one ``initial'' and one ``final'', carries two signature flips in the corner where the two hypersurfaces intersect. See figure \ref{fig:spacelike_boundary}.
       
At corners and/or flip surfaces, some standard tools of analytical mechanics cannot be taken for granted. This is related to the fact that the boundary-value problem for hyperbolic differential equations (i.e. for Lorentzian causality) is ill-defined. As a result, the usual formalism of boundary-data variations doesn't hold at all points of a closed boundary, but breaks down at the corners and flip surfaces. In particular, tangential and normal gradients on such surfaces cannot be treated as independent. This affects the counting of degrees of freedom, as well as the locality properties of action variations. One usually avoids these issues by keeping the boundary data on these surfaces fixed and not worrying about contributions that arise from them. This is quite natural when the surfaces are ``hidden'' at asymptotic infinity, where all the dynamical fields fall off. However, in the presence of gravity, this point of view becomes problematic. First, as mentioned above, asymptotic infinity may be physically inaccessible. Second, the boundary's metric and extrinsic curvature are now dynamical variables, and their values at infinity contribute to the action. This is the source of the divergence mentioned in footnote \ref{ftn:divergence}.

Thus, we are interested in studying actions in finite spacetime regions with an emphasis on the effects of corners and flip surfaces. The hope is to learn from this something about the degrees of freedom of quantum gravity in such regions. In \cite{NeimanOnShellActions}, this approach was followed for the case of closed null boundaries. From the perspective presented above, this is an extreme case, since the boundary is null not just on isolated surfaces, but everywhere. It was noticed in \cite{NeimanOnShellActions} that for GR in a null-bounded region, a careful evaluation of the action $S$ reveals an \emph{imaginary part}. This conclusion was extended to general closed boundaries in \cite{NeimanTheImaginaryPart}. The imaginary part arises from the action's boundary term, which requires analytical continuation in the vicinity of a flip surface. Its value closely resembles the black hole entropy formula, and can be written as:
\begin{align}
 \Im S = \frac{1}{4}\sum_{\mathrm{flips}}\sigma_{\mathrm{flip}} = \frac{1}{16G}\sum_{\mathrm{flips}}A_{\mathrm{flip}} \ . \label{eq:ImS}
\end{align}
Here, the sum is over flip surfaces, $A_{\mathrm{flip}}$ is the area of each surface, and $\sigma_{\mathrm{flip}}$ is the entropy functional $A/4G$ \cite{HawkingBlackHoleExplosions}. The calculation was also extended to Lovelock gravity, using the action with the appropriate boundary term \cite{MyersHigherDerivativeGravity}. This resulted again in an imaginary part $\Im S$, related in the same way as in \eqref{eq:ImS} to the appropriate entropy formula \cite{JacobsonBlackHoleEntropy}.
We note again the similarity to Gibbons' and Hawking's calculation \cite{GibbonsActionIntegralsAnd}, where black hole entropy was derived from an imaginary action. The motivation there, however, was somewhat different. A Wick rotation was performed to avoid the physical singularity in the black hole's interior. As a result, the calculation was restricted to stationary spacetimes, and did not involve finite regions.

\section{Boundary terms in first order general relativity} \label{sec:BoundaryTerm}

\subsection{Second order boundary term} \label{sec:BoundaryTerm:second}

\begin{figure}%
\centering%
\includegraphics[scale=0.75]{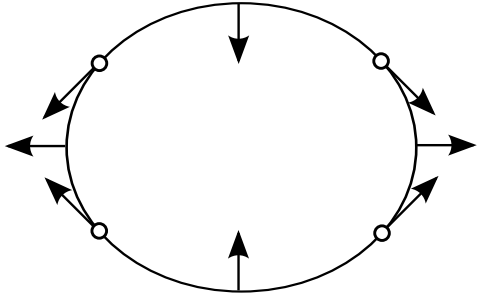} \\
\caption{A smooth closed boundary in Lorentzian spacetime. The arrows indicate the normal direction at various points. The normal's sign is chosen so that it has a positive scalar product with outgoing vectors. Empty circles denote ``signature flips'', where the normal becomes momentarily null.}
\label{fig:smooth_normal} 
\end{figure}

It is well-known that the Einstein-Hilbert action of GR must be supplemented with the York-Gibbons-Hawking boundary term \cite{YorkRoleOfConformal, GibbonsActionIntegralsAnd} to ensure a well-defined variational principle. The resulting action reads:
\begin{align}
 S_{\text{2nd-order}} = \frac{1}{16\pi G} \left( \int_\Omega \sqrt{-g}R\, d^4x + 2 \int_{\del\Omega} \sqrt{\frac{-h}{n\cdot n}} K\, d^3x \right) \ .
 \label{eq:EH_YGH}
\end{align}
Here, $\Omega$ is a spacetime region with boundary $\del\Omega$. $g_{\mu \nu}$ is the space-time metric, with signature $(-,+,+,+)$. $R = R^{\mu\nu}{}_{\mu\nu}$ is the Ricci scalar, with the sign convention $R^\mu{}_{\nu\rho\sigma}V^\nu = [\nabla_\rho,\nabla_\sigma]V^\mu$ for the Riemann tensor $R^\mu{}_{\nu\rho\sigma}$. $h_{ab}$ is the metric induced on $\del\Omega$. $K$ is the trace of the extrinsic curvature tensor $K_a^b = \nabla_a n^b$. The sign of the boundary normal $n^\mu$ is chosen so that the \emph{covector} $n_\mu$ is outgoing, i.e. so that $n^\mu$ has positive scalar product with outgoing vectors. See figure \ref{fig:smooth_normal}. The factor of $n\cdot n$ in the denominator makes \eqref{eq:EH_YGH} valid for both spacelike and timelike patches of $\partial \Omega$ and for arbitrary norm of $n^\mu$. Null regions are dealt with via an analytic continuation. $(\mu, \nu \ldots)$ are spacetime tensor indices, while $(a,b\ldots)$ are tensor indices on $\del\Omega$. 

Thanks to the boundary term, the action \eqref{eq:EH_YGH} contains only first derivatives. It is then stationary under the Einstein equations with $\delta h_{ab} = 0$ on $\del\Omega$. 
In addition, it's worth noting that a canonical analysis of the action \eqref{eq:EH_YGH} leads to the ADM energy and momentum \cite{ArnowittTheDynamicsOf} as the boundary terms of the Hamiltonian and spatial diffeomorphism constraints. A recent review of these results with extensions to $f(R)$ theories can be found in \cite{DyerBoundaryTermsVariational}. 

\subsection{The boundary term of the Palatini action} \label{sec:BoundaryTerm:Palatini} 

In addition to the second-order metric formulation, GR can also be described by the first-order Palatini action. There, one varies the co-vierbein $e^I_\mu$ and the SO$(1,3)$-connection $A_\mu^{IJ}$ independently. We use indices $(\mu,\nu,\dots)$ for spacetime coordinates, indices $(I,J,\dots)$ for the internal Minkowski space, and indices $(a,b,\dots)$ for coordinates on the boundary of our spatial region. Out of $e^I$ and $A_{IJ}$, we define the curvature $F_{IJ}$, the area-bivector 2-form $\Sigma_{IJ}$ and the densitized boundary normal $N_I$:
\begin{align}
 F_{IJ}(A) &:= dA_{IJ} + A_I{}^K\wedge A_{KJ} \\
 \Sigma^{IJ}(e) &:= \frac{1}{2} \epsilon^{IJKL} e_K \wedge e_L \\
 N_I(e) &:= \frac{1}{3!}\epsilon^{abc}\epsilon_{IJKL} e_a^J e_b^K e_c^L 
\end{align}
Our convention for the components of differential forms is $(U\wedge V)_{\mu\nu} = 2U_{[\mu}V_{\nu]}$. The spacetime Levi-Civita density is $\epsilon^{\mu\nu\rho\sigma}$, with inverse $\epsilon_{\mu\nu\rho\sigma}$. For the boundary, we have likewise $\epsilon^{abc}$ and $\epsilon_{abc}$. We define the relative sign of $\epsilon^{abc}$ and $\epsilon^{\mu\nu\rho\sigma}$ such that $\epsilon_{\mu abc}\epsilon^{abc}$ has a positive scalar product with outgoing vectors. The internal Levi-Civita tensor is $\epsilon^{IJKL}$ or $\epsilon_{IJKL}$; the two versions are related through raising and lowering with the flat metric $\eta_{IJ}$, and are \emph{minus} each other's inverses. We assume for now that the vierbein has a positive determinant $\det e = (1/4!)\epsilon^{\mu\nu\rho\sigma}\epsilon_{IJKL}e_\mu^I e_\nu^J e_\rho^K e_\sigma^L > 0$. This assumption can be thought of as $\epsilon_{\mu\nu\rho\sigma}$ and $\epsilon_{IJKL}$ ``having the same sign''. The case of negative $\det e$ will be discussed in section \ref{sec:BoundaryTerm:det_e}. With these sign conventions, the direction of the normal $N_I$ is as depicted in figure \ref{fig:smooth_normal}.

With these ingredients in place, we argue for the following form of the Palatini action \cite{ObukhovThePalatiniPrinciple,WielandComplexAshtekarVariables,BianchiHorizonEnergyAs}:
\begin{align}
 \begin{split}
   S &= \frac{1}{16\pi G} \left(\int_\Omega \Sigma^{IJ} \wedge F_{IJ}
     - \int_{\del\Omega} \Sigma^{IJ} \wedge \left(A_{IJ} - \frac{2N_I dN_J}{N\cdot N}\right) \right) \\
    &= \frac{1}{16\pi G} \left(\int_\Omega \Sigma^{IJ} \wedge F_{IJ}
     + \int_{\del\Omega} \frac{2}{N\cdot N}\, \Sigma^{IJ} \wedge N_I (dN_J + A_J{}^K N_K) \right) \ .
 \end{split} \label{eq:Palatini_N} 
\end{align}
The two expressions are equivalent due to the fact that $\Sigma^{IJ}$ has one of its internal indices in the direction of $N_I$. The action is more commonly written in terms of a unit normal. The densitized normal has two advantages. First, it is a simple polynomial function of the $e^I$. Second, it remains non-singular when the boundary becomes null, while the unit normal diverges. It is thus better suited for describing flip surfaces. Alternatively, one can write the action in terms of an undensitized, \emph{non-unit} normal $n_I$:
\begin{align}
  S &= \frac{1}{16\pi G} \left(\int_\Omega \Sigma^{IJ} \wedge F_{IJ}
    - \int_{\del\Omega} \Sigma^{IJ} \wedge \left(A_{IJ} - \frac{2n_I dn_J}{n\cdot n}\right) \right) \label{eq:Palatini_simple} \\
   &= \frac{1}{16\pi G} \left(\int_\Omega \Sigma^{IJ} \wedge F_{IJ}
    + \int_{\del\Omega} \frac{2}{n\cdot n}\, \Sigma^{IJ} \wedge n_I Dn_J \right) \ , \label{eq:Palatini_invariant}
\end{align}
where $D n_J = dn_J + A_J{}^K n_K$ is the covariant derivative of $n_I$. In this formulation, rescalings of $n_I$ constitute a gauge freedom. Using $n_I$ is perhaps more intuitive, and we will do so in our analysis of the action. The relation between $N_I$ and $n_I$ is:
\begin{align}
 N_I = \sqrt{\frac{-h}{n\cdot n}}\,n_I \ , \label{eq:N_n}
\end{align}
where $h$ is the determinant of the boundary metric $h_{ab} = e_{aI} e_b^I$.

When written in the form \eqref{eq:Palatini_invariant}, the action is manifestly gauge-invariant. It is also clear that the boundary term agrees on-shell with the York-Gibbons-Hawking boundary term from \eqref{eq:EH_YGH}. More precisely, the boundary terms agree when $A_{IJ}$ coincides with the spin connection of $e^I$. This ``half-shell'' condition results from varying the action with respect to $A_{IJ}$. It is well-known that the bulk terms also agree under the same condition. We conclude that our Palatini action coincides on-shell with the second-order action. As we'll see in more detail in section \ref{sec:BoundaryTerm:corners}, this implies that they also have the same imaginary part.      

Let us examine the ingredients of the Palatini boundary term in more detail. First, consider the action with no boundary term at all. Its variation $\delta S$ on a solution is a boundary integral of $\Sigma^{IJ}\wedge\delta A_{IJ}$.  Thus, the variational principle contains boundary conditions on $\delta A^{IJ}$ rather than on $\delta e^I$. Now, the $\Sigma\wedge A$ piece of the action's boundary term turns $\delta S$ into a boundary integral of $\delta\Sigma^{IJ}\wedge A_{IJ}$. The appropriate boundary conditions become $\delta e_a^I = 0$, with $\delta A_{IJ}$ arbitrary. As we will see in section \ref{sec:BoundaryTerm:corners}, the $\Sigma^{IJ}\wedge n_I dn_J$ piece of the boundary term allows for slightly weaker boundary conditions, in agreement with the second-order variational principle.

In the literature \cite{AshtekarAsymptoticsAndHamiltoniansFour,CorichiSurfaceTermsAsymptotics}, one can find boundary terms that consist solely of the $\Sigma^{IJ}\wedge A_{IJ}$ piece. As discussed in \cite{AshtekarAsymptoticsAndHamiltoniansFour}, if one adopts the time gauge $n^I = (1,0,0,0)$ on the boundary, then the $\Sigma^{IJ}\wedge A_{IJ}$ piece is invariant under the residual gauge transformations. One can then argue that the action is sufficiently gauge-invariant without an $\Sigma^{IJ}\wedge n_I dn_J$ piece in the boundary term. However, in our context of closed boundaries, the question is irrelevant: time gauge is just not available globally. In Lorentzian spacetime, this is obvious due to changes in the normal's signature and/or time-orientation. In fact, time gauge is not available in the Euclidean either, for topological reasons: though one can choose a frame on the boundary where $n^I$ is constant, one cannot extend it smoothly into the region's bulk. For a simple example, consider a circular boundary in a Euclidean plane.

We conclude, then, that the $\Sigma^{IJ}\wedge n_I dn_J$ piece of the boundary term is essential for a gauge-invariant action. This in turn implies that a sensible Palatini action must agree on-half-shell with the second-order action \eqref{eq:EH_YGH}. In the next subsection, we will see how the $\Sigma^{IJ}\wedge n_I dn_J$ piece is responsible for corner contributions and for the action's imaginary part.   

\subsection{Corner contributions and the imaginary part of the Palatini action} \label{sec:BoundaryTerm:corners}

The $\Sigma^{IJ}\wedge n_I dn_J$ piece of the Palatini boundary term \eqref{eq:Palatini_simple} restores gauge invariance after the introduction of the $\Sigma^{IJ}\wedge A_{IJ}$ term. As a result, it allows for weaker boundary conditions in the variational principle. Due to the action's gauge invariance, we can now allow boundary variations of $e_a^I$, as long as the intrinsic metric variation $h_{ab} = e_{aI} e_b^I$ stays fixed ($\delta A_a^{IJ}$ remains arbitrary). On smooth patches of the boundary, this new condition is gauge-equivalent to simply keeping $e_a^I$ fixed. Indeed, the three Minkowski vectors $e_a^I$ are determined by their scalar products $h_{ab}$, up to rotations in the internal space. 

At corners, where the boundary turns abruptly by a nonzero angle, the boundary conditions $\delta e_a^I = 0$ and $\delta h_{ab} = 0$ are {\it no longer} gauge-equivalent. Fixing $e_a^I$ at corners is a slightly too strong condition: in addition to the intrinsic metric, it fixes also the corner angle, which should properly be part of the extrinsic curvature. Fixing $h_{ab} = e_{aI} e_b^I$ instead leaves the corner angles free to vary, and is equivalent to the standard boundary conditions of second-order gravity. The merit of this choice of boundary conditions is that it keeps the fixed data on the boundary to a minimum. Given free variations of the corner angles, one is forced by the variational principle \cite{HaywardGravitationalActionFor} to take into account corner contributions \cite{HartleBoundaryTermsIn} to the action's boundary term. Corner contributions will be crucial in section \ref{sec:SpinFoam}, since the 4-simplex action studied there consists of nothing else.  

The $\Sigma^{IJ}\wedge n_I dn_J$ piece of the boundary term not only leads to a variational principle that necessitates corner contributions. It also contains the corner contributions themselves. To begin with, consider a flat solution. There, one can choose coordinates and a gauge frame such that $e_\mu^I = \mbox{const}$ and $A_\mu^{IJ} = 0$. In this frame, the bulk term and the $\Sigma^{IJ}\wedge A_{IJ}$ boundary term in \eqref{eq:Palatini_simple} both vanish, for any chosen region. On the other hand, the $\Sigma^{IJ}\wedge n_I dn_J$ term does not vanish. In fact, it precisely captures the York-Gibbons-Hawking contribution from the boundary's extrinsic curvature. In particular, for regions $\Omega$ that are effectively 1+1-dimensional (say, the other two dimensions form a flat torus), the on-shell action reads:
\begin{align}
 S = \frac{1}{8\pi G}\int_{\del\Omega} \Sigma^{IJ}\wedge \frac{n_I dn_J}{n\cdot n} 
  = \frac{1}{8\pi G}\int_{\del\Omega} \sqrt{\frac{-h}{n\cdot n}} K\, d^3x = \frac{A}{8\pi G}\int d\alpha \ . \label{eq:2d}
\end{align}
Here, $d\alpha$ is the normal's rotation angle (actually, boost parameter) in the 1+1d space, while $A$ is the area of the two transverse dimensions. To see this, it might be helpful to write $K$ as $e^{aI} \del_a n_I$ (recall our flat frame choice), where $e^{aI} := h^{ab} e_b^I$.

The example of an effectively 1+1d region in flat spacetime is less artificial than it may seem. In particular, it applies in the infinitesimal neighborhood of a corner surface. This is because the boundary's extrinsic curvature in the orthogonal 1+1d plane is arbitrarily larger than the gradients of $e^I$ or the components of $A_{IJ}$ (assuming a smooth gauge frame). Therefore, the flat 1+1d discussion captures the fate of the corner contributions \cite{HartleBoundaryTermsIn,HaywardGravitationalActionFor} to the York-Gibbons-Hawking boundary term. We see that the corner contributions are due entirely to the $\Sigma^{IJ}\wedge n_I dn_J$ piece of the boundary term.

Now, as explained in \cite{NeimanTheImaginaryPart}, the signature-flip surfaces on $\del\Omega$ behave like corners, whether or not they coincide with corners in the topological sense. The ``corner contribution'' from these surfaces is responsible for the imaginary part \eqref{eq:ImS} of the on-shell action, and is due to the imaginary part of the ``corner angle''. We conclude that the imaginary part $\Im S$ is again due to the $\Sigma^{IJ}\wedge n_I dn_J$ boundary term.

In more detail, at a signature-flip surface, when the boundary becomes momentarily null, the $n\cdot n$ in the denominator in \eqref{eq:Palatini_simple}-\eqref{eq:Palatini_invariant} vanishes (assuming the extent of $n_I$ is kept finite). This leads to a divergence in the boundary term's integrand. This is the same situation that was studied for the York-Gibbons-Hawking boundary term in \cite{NeimanTheImaginaryPart}. As explained there, the divergence in the boundary term is simply the divergence in the angle integral \eqref{eq:2d} as the normal approaches a null direction. The regularization of the angle integral yields an imaginary part $\pm \pi/2$, depending on a choice of analytical continuation. In the action, this translates into the imaginary part \eqref{eq:ImS}. In terms of the action's bounadry term, the analytical continuation involved is equivalent to regularizing the denominator as $n\cdot n \rightarrow n\cdot n \pm i\epsilon$. The sign of $i\epsilon$, and thus of $\Im S$, is chosen so that $\Im S$ is non-negative. This means that transition amplitudes $e^{iS}\sim e^{-\Im S}$ are exponentially damped rather than exponentially exploding. As discussed in \cite{NeimanTheImaginaryPart}, this damping seems consistent with the multiplicity of states associated with gravitational entropy. 

\subsection{Holst action} \label{sec:BoundaryTerm:Holst}

The Palatini action can be supplemented by the so-called Holst modification \cite{HojmanParityViolationIn,HolstBarberosHamiltonianDerived}, without changing the equations of motion. Together with the appropriate boundary term, the action then reads:
\begin{align}
 \begin{split}
   S ={}& \frac{1}{16\pi G}\left( \int_\Omega \Sigma^{IJ} \wedge F_{IJ} 
     - \int_{\del\Omega} \Sigma^{IJ} \wedge \left(A_{IJ} - \frac{2N_I dN_J}{N\cdot N}\right) \right) \\
    &+ \frac{1}{16\pi G\gamma} \left(\int_\Omega e^I\wedge e^J \wedge F_{IJ} 
      - \int_{\del\Omega}e^I\wedge(e^J\wedge A_{IJ} - de_I)\right) \ , 
 \end{split} \label{eq:Holst_simple} 
\end{align}
where $\gamma$ is a real nonzero constant. The first line is the Palatini action from section \ref{sec:BoundaryTerm:Palatini}, while the second line describes the Holst modification. The above expression makes clear the dependence on the fundamental variables. To make gauge invariance manifest, we can rewrite eq. \eqref{eq:Holst_simple} as:
\begin{align}
 \begin{split}
   S ={}& \frac{1}{16\pi G}\left( \int_\Omega \Sigma^{IJ} \wedge F_{IJ} 
     + \int_{\del\Omega} \frac{2}{n\cdot n}\, \Sigma^{IJ} \wedge n_I Dn_J \right) \\
    &+ \frac{1}{16\pi G\gamma} \left(\int_\Omega e^I\wedge e^J \wedge F_{IJ} 
      + \int_{\del\Omega}e^I\wedge De_I \right) \ , 
 \end{split} \label{eq:Holst_invariant}
\end{align}
using a non-densitized normal $n_I$. 

The $e^I\wedge e^J\wedge A_{IJ}$ term in the boundary integral in \eqref{eq:Holst_simple} performs the same function as the $\Sigma^{IJ}\wedge A_{IJ}$ term in the Palatini action - it ensures that the variational principle fixes metric data on the boundary rather than connection data. The $e^I\wedge de_I$ term then performs the same function as the $\Sigma^{IJ}\wedge N_I dN_J$ term in the Palatini action - it restores full gauge invariance, and allows for a variational principle where only the intrinsic boundary metric $h_{ab} = e_{aI}e_b^I$ is held fixed. Unlike the $\Sigma^{IJ}\wedge N_I dN_J$ term, the $e^I\wedge de_I$ term has no denominators, and is polynomial in $e^I$. It doesn't diverge at corners or flip surfaces, and thus doesn't require analytical continuation.

It is easy to see that the on-shell action is unaffected by the Holst modification. Both the bulk and boundary integrands in the second line of \eqref{eq:Holst_invariant} vanish, due to the ``half-shell'' identities:
\begin{align}
 De_I = 0 \quad;\quad e^J\wedge F_{IJ} = 0 \ . \label{eq:identities}
\end{align}
The first equation in \eqref{eq:identities} is the defining property of the spin-connection, while the second is the Bianchi identity. We conclude that on-shell, the Holst action \eqref{eq:Holst_simple} coincides with the modified Palatini action \eqref{eq:Palatini_N}, and therefore also with the second-order action \eqref{eq:EH_YGH}.

The Holst modification does have an effect on the Hamiltonian formalism. A canonical analysis of the Holst action \eqref{eq:Palatini_N} leads to the Ashtekar-Barbero variables \cite{AshtekarNewVariablesFor, BarberoRealAshtekarVariables}, with $\gamma$ as the Barbero-Immirzi parameter. From a Hamiltonian point of view, different values of $\gamma$ amount to canonical transformations \cite{BarberoRealAshtekarVariables}. 

While $\gamma$ does not affect the physics at the classical level, it becomes important in loop quantum gravity \cite{RovelliQuantumGravity, ThiemannModernCanonicalQuantum}, where it appears as a quantization ambiguity. Here, the canonical transformation describing a change of $\gamma$ cannot be implemented on the holonomy-flux algebra (see, e.g., \cite{ThiemannModernCanonicalQuantum}), which leads to unitarily inequivalent quantum theories \cite{RovelliTheImmirziParameter}. $\gamma$ can thus enter the spectra of physical observables \cite{ImmirziQuantumGravityAnd}, and predictions of the quantum theory can depend on it. 

It has been debated in the literature \cite{MercuriFermionsInThe, MercuriAPossibleTopological, DateTopologicalInterpretationOf} whether one should use the Nieh-Yan topological density $d(e^I \wedge De_I)$ instead of the Holst modification as the classical starting point for loop quantum gravity. The argument in favor of the Nieh-Yan density is that because it's truly topological, it does not affect the fermion coupling in the Lagrangian (see however \cite{AlexandrovTheImmirziParameter}). As a result, the theory with fermions remains insensitive to the Barbero-Immirzi parameter at the classical level. However, our analysis above shows that to have a variational principle with arbitrary $\delta A_{IJ}$ on the boundary, one must use a combination of the Holst and Nieh-Yan terms given in \eqref{eq:Holst_invariant}. This subtlety will also be important when discussing the running of the Barbero-Immirzi parameter, since prefactors of topological terms, e.g. the $\theta$-angle in QCD, are not expected to run in perturbation theory.

As a side remark, the action \eqref{eq:Holst_simple} can be written in a manifestly dual way as:\footnote{We thank Andreas Thurn for this observation  (private communication).}
\begin{align}
 \begin{split}
   S = \quad\ &\frac{1}{16\pi G} \int_{\Omega} \left( \frac{1}{2} \epsilon^{IJKL} e_K \wedge e_L + \frac{1}{\gamma} e^I \wedge e^J\right) \wedge F_{IJ} \\
    -& \frac{1}{16\pi G} \int_{\del\Omega} \left( \frac{1}{2} \epsilon^{IJKL} e_K \wedge e_L + \frac{1}{\gamma} e^I \wedge e^J \right) 
    \wedge \left(A_{IJ} - \Gamma^\text{H}_{IJ} \right) \ ,
 \end{split} \label{eq:HolstActionManifestlyDual}
\end{align}
where $\Gamma^\text{H}_{IJ}$ is a functional of $e_a^I$ known as the Peldan hybrid connection \cite{PeldanActionsForGravity}.

\subsection{The first-order action for negative $\det e$} \label{sec:BoundaryTerm:det_e}

In the above calculations, we've been assuming $\det e > 0$. Let us now consider the action's behavior when both signs of $\det e$ are allowed. This will be relevant to the spinfoam results in section \ref{sec:SpinFoam}. We build here on the discussion in \cite{RovelliDiscreteSymmetriesIn}. Note that the sign of $\det e$ encodes the orientation of the internal space with respect to the spacetime manifold. Thus, changing this sign can be viewed as a parity or time-reversal transformation.

When $\det e$ changes sign, both the bulk and boundary terms in the Palatini action \eqref{eq:Palatini_N} acquire a sign factor. Thus, for $\det e < 0$, the real part of \eqref{eq:Palatini_N} evaluates on-shell to \emph{minus} the real part of the second-order action \eqref{eq:EH_YGH}. As for the imaginary part, recall that its sign is determined by a choice between two complex-conjugate analytical continuations of the boundary normal's angle \cite{NeimanOnShellActions,NeimanTheImaginaryPart}. The requirement that quantum amplitudes $e^{iS}\sim e^{-\Im S}$ do not explode exponentially forces us to choose a positive $\Im S$, regardless of the sign of $\det e$. Thus, the relative sign between the real and imaginary parts of the boundary term, i.e. the choice of analytical continuation, must be different for $\det e > 0$ and for $\det e < 0$.

Alternatively, one can multiply the Palatini action \eqref{eq:Palatini_N} by a factor of $\sign(\det e)$ (in the classical theory, this sign will be constant everywhere, assuming that $\det e$ doesn't vanish):
\begin{align}
 S' = \frac{\sign(\det e)}{16\pi G} \left(\int_\Omega \Sigma^{IJ} \wedge F_{IJ}(A) 
    - \int_{\del\Omega} \Sigma^{IJ} \wedge \left(A_{IJ} - \frac{2N_I dN_J}{N\cdot N}\right) \right) \ . \label{eq:S'}
\end{align}
The action \eqref{eq:S'} is invariant under changes to $\sign(\det e)$, and its real part always evaluates on-shell to the second-order action \eqref{eq:EH_YGH}. As a result, the same analytical continuation gives a positive $\Im S'$ for both signs of $\det e$. The price we pay is that the action functional \eqref{eq:S'} is no longer fully analytical in the co-vierbein. The sign function in \eqref{eq:S'} can be analytically continued either from positive $\det e$ or from negative $\det e$. The two continuations give the constant functions $\sign(\det e)=1$ and $\sign(\det e)=-1$, thus forming two separate analytical domains. We see that there is a tradeoff in terms of analyticity. For the action \eqref{eq:Palatini_N}-\eqref{eq:Palatini_invariant}, different signs of $\det e$ require different analytical continuations for the normal's angle. On the other hand, with the action \eqref{eq:S'}, we must contend with the two separate analytical domains of the sign function $\sign(\det e)$.

The appearance of two analytical domains is related to the 4-volume density $\sqrt{-g}$ in the second-order action\footnote{We are focusing here on the bulk term. A similar argument can be made for the boundary term.}. In terms of the vierbein, we have $g = -(\det e)^2$, and therefore:
\begin{align}
 \sqrt{-g} = \sqrt{(\det e)^2} = \sign(\det e)\cdot\det e \ . \label{eq:g_e}
\end{align}
While \eqref{eq:g_e} shows where the non-analytical factor of $\sign(\det e)$ comes from, it is instructive to consider the function $\sqrt{(\det e)^2}$ in its own right. The square-root function has a branch cut induced by the choice $\sqrt{-1} = \pm i$. This means that a half-line, e.g. $e^{i\theta} \mathbb{R}^+$ for some angle $\theta\neq\pi$, must be removed from its domain. For the function $\sqrt{(\det e)^2}$, this implies removing from the complex $\det e$ plane a whole line containing $\det e = 0$, e.g. $e^{i\theta/2} \mathbb{R}$. This line splits the complex plane into two separate domains, one containing the positive half of the real line, and the other containing the negative half.

Similar considerations apply to the Holst action \eqref{eq:Holst_simple}-\eqref{eq:Holst_invariant}. Under a change in $\sign(\det e)$, the first line in \eqref{eq:Holst_simple} changes sign, while the second line (the $1/\gamma$ term) is invariant. Since the second line vanishes on-shell, the comparison with the second-order action and the choice of analytical continuation for the angle play out as before. The analogue of the action \eqref{eq:S'} reads:
\begin{align}
 \begin{split}
   S' ={}& \frac{\sign(\det e)}{16\pi G} \left(\int_\Omega \Sigma^{IJ} \wedge F_{IJ}(A) 
     - \int_{\del\Omega} \Sigma^{IJ} \wedge \left(A_{IJ} - \frac{2N_I dN_J}{N\cdot N}\right)\right) \\
    &+ \frac{1}{16\pi G\gamma} \left(\int_\Omega e^I\wedge e^J \wedge F_{IJ}(A) 
      - \int_{\del\Omega}e^I\wedge(e^J\wedge A_{IJ} - de_I)\right) \ .
 \end{split} \label{eq:HolstS'} 
\end{align}
The action \eqref{eq:HolstS'} is invariant under changes to $\sign(\det e)$ and reduces on-shell to the second-order action \eqref{eq:EH_YGH} for both signs of $\det e$. As in \eqref{eq:S'}, the price is the appearance of an explicit sign function, which splits the complex plane into two analytical domains. Such an action was proposed in \cite{RovelliDiscreteSymmetriesIn} as the basis of a modified spinfoam model that would be invariant under parity and time reversal. Similar proposals were made in \cite{MikovicEffectiveActionAnd,MikovicEffectiveActionFor,EngleASpinFoam,EngleAProposedProper}, with the purpose of improving the large-spin behavior. In section \ref{sec:SpinFoam}, we will discuss the original EPRL/FK spinfoams based on the action \eqref{eq:Holst_simple}-\eqref{eq:Holst_invariant}, since this is the model that has been most extensively studied in the Lorentzian.

\section{The imaginary action in spinfoam asymptotics} \label{sec:SpinFoam}

\subsection{Introduction}

We have seen that various formulations of the GR action possess an imaginary part \eqref{eq:ImS} when supplemented with a gauge-invariant boundary term. In particular, this is true for the Holst action, which forms a heuristic basis for the ``new'' (EPRL/FK) spinfoam model \cite{FreidelANewSpin,EngleLoopQuantumGravity}. It is therefore interesting to see whether the imaginary action \eqref{eq:ImS} makes an appearance in the new spinfoams. As the signature flips that lead to an imaginary action are a crucially Lorentzian feature, we focus on the Lorentzian spinfoam model.    

Since the model is based on an action formula only heuristically, we must look at its output, i.e. at the transition amplitudes. In a classical limit, the logarithm of a quantum amplitude (divided by $i$) acquires the meaning of an action: $\ln A \rightarrow iS$. In the spinfoam model, a simplified semiclassical analysis can be performed by considering geometries with very large discrete elements, i.e. with spins $j\gg 1$ on the spinfoam faces. One can then consider coherent boundary states peaked on specific values of the intrinsic metric and extrinsic curvature (which are non-commuting variables) \cite{BianchiLorentzianSpinfoamPropagator}. A useful middle ground are so-called ``semi-coherent'' boundary states, which were studied in \cite{BarrettLorentzianSpinFoam}. These are similar to fully-coherent states, but have definite values for the areas, i.e. for the spins $j$.

\subsection{The 4-simplex spinfoam action for large spins}

\begin{figure}%
\centering%
\includegraphics[scale=0.75]{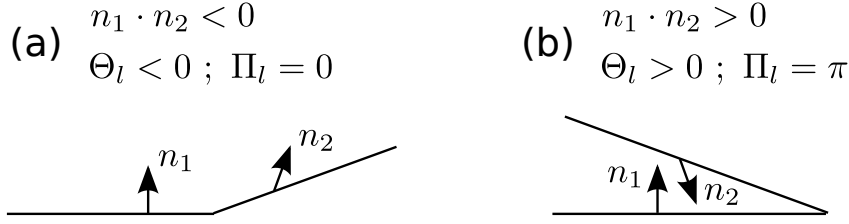} \\
\caption{Two types of corners between spacelike tetrahedra: thick (a) and thin (b). The arrows denote ingoing unit normals. These correspond to outgoing covectors, i.e. have a positive scalar product with outgoing vectors. For each type of corner, the content of eqs. \eqref{eq:Theta}-\eqref{eq:Pi} is summarized.}
\label{fig:thick_thin} 
\end{figure}

In \cite{BarrettLorentzianSpinFoam}, Barrett et.al. studied the spinfoam vertex amplitude $f_4$ for a large-spin semi-coherent boundary state describing a flat 4-simplex - the simplest non-trivial polytope in spacetime. The boundary is composed of five spacelike tetrahedra, intersecting at ten triangles. For boundary data that yields a consistent 4-simplex geometry, the authors of \cite{BarrettLorentzianSpinFoam} reduced the vertex amplitude to a sum of two saddle-point contributions. The two saddle points describe two configurations of the 4-simplex related by parity (or time reversal). A fully coherent boundary state that includes a smearing over spins picks out one of these saddle points \cite{BianchiLorentzianSpinfoamPropagator}. Taking this into account, we write the result of \cite{BarrettLorentzianSpinFoam} as:
\begin{align}
 f_4 \rightarrow (-1)^\chi N_\pm e^{iS_\pm} \ , \label{eq:SimplexAmplitude}
\end{align}
where the sign in the subscripts distinguishes the two saddle points. The sign factor $(-1)^\chi$ arises from the combinatorics of the boundary graph. The coefficients $N_\pm$ are ``weak'' functions of the boundary data, i.e. their contribution to $\ln f_4$ is negligible (in \cite{NeimanParityAndReality}, numerical evidence was given to the effect that $N_+$ and $N_-$ are complex conjugates). Thus, we have $\ln f_4\rightarrow iS_\pm$, so that the quantity $S_\pm$ has the usual meaning of action. On this point, we deviated from the notation in \cite{BarrettLorentzianSpinFoam}: there, the exponent in \eqref{eq:SimplexAmplitude} is itself called an ``action''. This means that our action differs from that in \cite{BarrettLorentzianSpinFoam} by a factor of $i$. With this difference in mind, the result of \cite{BarrettLorentzianSpinFoam} for $S_\pm$ reads:
\begin{align}
 S_\pm = \sum_l j_l\left(\pm\gamma\Theta_l - \Pi_l\right) \ . \label{eq:SpinfoamS_raw}
\end{align}
Here, we kept the second term, which in \cite{BarrettLorentzianSpinFoam} was taken outside the exponent $e^{iS_\pm}$. The reason for this will become clear below.
$\gamma$ in \eqref{eq:SpinfoamS_raw} is the Barbero-Immirzi parameter. The sum is over the links in the boundary spin-network, or, equivalently, the faces of the spinfoam. Geometrically, they correspond to the triangular 2d faces of the 4-simplex. Each of these faces is a ``corner'', in the sense that two boundary hypersurfaces (in this case, tetrahedral hyperfaces) intersect there at an angle. The $j_l$ are spin variables which are proportional to the triangle areas $A_l$. We will elaborate on the precise nature of this proportionality below. Eq. \eqref{eq:SpinfoamS_raw} is derived in the limit $j_l\gg 1$. The $\Theta_l$ are the boost angles between the pairs of tetrahedra intersecting at each triangle. Let $n_1^\mu$ and $n_2^\mu$ be the timelike unit normals to the pair of tetrahedra intersecting at the triangle $l$. We choose $n_1^\mu$ and $n_2^\mu$ to be both ingoing with respect to the 4-simplex (taking into account the timelike signature, this gives the normals a positive scalar product with outgoing vectors \cite{NeimanTheImaginaryPart}). The angle $\Theta_l$ is then defined by:
\begin{align}
 \Theta_l = &\left\{
  \begin{array}{ll}
    -\acosh(-n_1\cdot n_2) & \quad n_1\cdot n_2 < 0 \quad \mbox{(thick corner)} \\
    +\acosh(+n_1\cdot n_2) & \quad n_1\cdot n_2 > 0 \quad \mbox{(thin corner)}                
  \end{array} \right. \ . \label{eq:Theta}
\end{align}
Here, a ``thick corner'' is one where both normals $n^\mu_1,n^\mu_2$ have the same time orientation, while a ``thin corner'' is one where the time orientations are opposite. See figure \ref{fig:thick_thin}. Finally, the $\Pi_l$ in \eqref{eq:SpinfoamS_raw} are defined as:
\begin{align}
 \Pi_l = &\left\{
  \begin{array}{ll}
    0 & \quad \mbox{thick corner} \\
    \pi & \quad \mbox{thin corner}                
  \end{array} \right. \ . \label{eq:Pi}
\end{align}

Let us now return to the proportionality between the spins $j_l$ in \eqref{eq:SpinfoamS_raw} and the triangle areas $A_l$. It is given by:
\begin{align}
 A_l = 8\pi G\sign(\gamma)\gamma j_l \ , \label{eq:A}
\end{align}
where we keep in mind that $\gamma$ can have either sign, while the areas $A_l$ must be positive. We refrain from replacing $\sign(\gamma)\gamma \rightarrow |\gamma|$, since that would interfere with the analytical continuation below. Substituting \eqref{eq:A} into \eqref{eq:SpinfoamS_raw}, the action becomes:
\begin{align}
 S_\pm = \frac{\sign(\gamma)}{8\pi G}\sum_l A_l\left(\pm\Theta_l - \frac{\Pi_l}{\gamma}\right) \ . \label{eq:SpinfoamS}
\end{align}
 
\subsection{The classical action for a 4-simplex}

For comparison, let us now find the classical GR action for a flat 4-simplex, including the imaginary part described in \cite{NeimanOnShellActions,NeimanTheImaginaryPart}. Having shown that the different classical actions agree on-shell up to signs, we will use for this purpose the second-order action \eqref{eq:EH_YGH}: 
\begin{align}
 S_{\text{2nd-order}} = \frac{1}{16\pi G}\int_\Omega \sqrt{-g}\, R\, d^4x 
  + \frac{1}{8\pi G}\int_{\del\Omega} \sqrt{\frac{-h}{n\cdot n}}\,K\, d^{3}x \ . \label{eq:EH_again}
\end{align}
While the Holst action \eqref{eq:Holst_simple} is more closely related to the spinfoam model, we cannot use it directly in our comparison. This is because of the sensitivity of \eqref{eq:Holst_simple} to the sign of $\det e$, which has no direct counterpart in the spinfoam variables. As we'll discuss in section \ref{sec:SpinFoam:compare}, one can say that the spinfoam result \eqref{eq:SpinfoamS} retains the dependence on $\sign(\det e)$ through the $\pm$ sign that distinguishes the two saddle points.  

\begin{figure}%
\centering%
\includegraphics[scale=0.75]{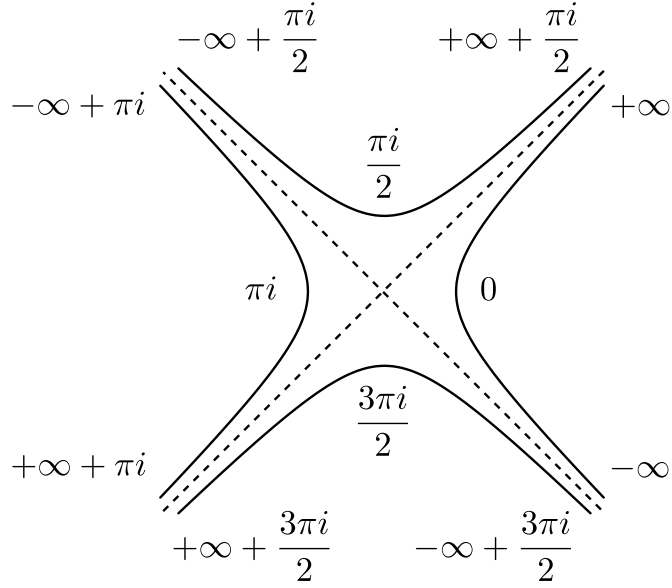} \\
\caption{An assignment of boost angles to points in a Lorentzian plane. The points represent values of the boundary normal $n^\mu$ in the 1+1d plane transverse to a corner (2d face) of the 4-simplex. The angles are defined up to integer multiples of $2\pi i$.}
\label{fig:angles_plane} 
\end{figure}

Since we are in flat spacetime, the bulk term in \eqref{eq:EH_again} vanishes. The boundary's extrinsic curvature, and with it the York-Gibbons-Hawking boundary term, is concentrated on the 2d triangular corners. We evaluate these corner terms according to the recipe in \cite{NeimanTheImaginaryPart}. At each corner, we consider the normals to the two intersecting tetrahedra, depicted in figure \ref{fig:thick_thin}. Each normal can be assigned an angle (actually, a boost parameter) in the 1+1d plane orthogonal to the corner triangle. To cover more than one quadrant of the Lorentzian plane, the angles are necessarily complex. The absolute value of the angles' real part is described as usual by hyperbolic functions. The sign and the imaginary part are depicted in figure \ref{fig:angles_plane}. The corner contributions to the action are then given by:
\begin{align}
 S_{\text{2nd-order}} = \frac{1}{8\pi G}\sum_l\alpha_l A_l \ , \label{eq:ClassicalS_raw}
\end{align}
where $A_l$ is the corner's area, and $\alpha_l$ is the angle difference between the two normals as one travels counter-clockwise in figure \ref{fig:thick_thin}. Comparing with eqs. \eqref{eq:Theta}-\eqref{eq:Pi}, we find that the corner angles $\alpha_l$ can be written as:
\begin{align}
 \alpha_l = -\Theta_l + i\Pi_l \ .
\end{align}
This brings the classical action \eqref{eq:ClassicalS_raw} to the form:
\begin{align}
 S_{\text{2nd-order}} = \frac{1}{8\pi G}\sum_l A_l(-\Theta_l + i\Pi_l) \ . \label{eq:ClassicalS}
\end{align}

\subsection{Comparison and analytical continuation in $\gamma$} \label{sec:SpinFoam:compare}

Let us now compare the spinfoam ``action'' \eqref{eq:SpinfoamS} with the classical 4-simplex action \eqref{eq:ClassicalS}. The spinfoam formula \eqref{eq:SpinfoamS} was derived for real values of $\gamma$. For such values, $S_\pm$ are also real, and therefore cannot reproduce the complex result \eqref{eq:ClassicalS}. For real $\gamma$, the $\Pi_l$ terms sum up to an integer multiple of $\pi$ \cite{BarrettLorentzianSpinFoam}, yielding a simple sign factor in the amplitude $e^{iS_\pm}$. As a result, in \cite{BarrettLorentzianSpinFoam}, the $\Pi_l$ terms were separated from the action. 

However, if we analytically continue \eqref{eq:SpinfoamS} to complex $\gamma$, the $\Pi_l$ terms can be made to reproduce the imaginary part of the classical action. The non-trivial part of the analytical continuation is the factor of $\sign(\gamma)$ in \eqref{eq:SpinfoamS}. This must be continued to a constant function, whose value depends on whether we start from $\gamma>0$ or from $\gamma<0$. Thus, we must properly speak of two separate analytical continuations\footnote{As with $\det e$ in section \ref{sec:BoundaryTerm:det_e}, it is instructive to view the two analytical domains as arising from the function $\sqrt{\gamma^2}$. This is because the more basic operator in loop quantum gravity is not the area $A_l\sim\sign(\gamma)\gamma$, but its square $A_l^2\sim\gamma^2$.}: one from $\gamma>0$ with $\sign(\gamma) = 1$, and one from $\gamma<0$ with $\sign(\gamma) = -1$. With this in mind, let us analytically continue the spinfoam action $S_\pm$ to $\gamma = \sign(\gamma) i$. Depending on the original value of $\sign(\gamma)$, this corresponds heuristically to a self-dual or anti-self-dual connection. Making the substitution $\gamma \rightarrow \sign(\gamma)i$ in \eqref{eq:SpinfoamS}, we get:
\begin{align}
 S_\pm \rightarrow \mp\sign(\gamma)\Re S_{\text{2nd-order}} + i \Im S_{\text{2nd-order}} \ , \label{eq:SpinfoamS_complex}
\end{align}
where $S_{\text{2nd-order}}$ is the classical second-order action \eqref{eq:ClassicalS}. 
Note that we keep the area fixed in the analytic continuation. It would not make sense to heuristically call $8 \pi G i j_l$ an area element, that is to continue $\gamma$ and keep the $j_l$ fixed, because it results in a complex number as area eigenvalue. For real $\gamma$, the area operator is self-adjoint and thus has real spectrum, a property one would have to retain in a proper quantization for complex $\gamma$.

The result \eqref{eq:SpinfoamS_complex} has the following properties:
\begin{enumerate}
 \item The imaginary part coincides with that of the classical action. In particular, it is positive by construction for both saddle points, giving exponentially suppressed amplitudes $e^{iS_\pm}$.
 \item The real part coincides with that of the second-order classical action \eqref{eq:EH_YGH}, up to sign. This sign is opposite for the two saddle points.
 \item The amplitudes $e^{iS_+}$ and $e^{iS_-}$ for the two saddle points are complex-conjugate to each other. This was also the case for real $\gamma$.
 \item Switching the sign of $\gamma$ corresponds to complex-conjugating $iS_\pm$. Since $iS_+$ and $iS_-$ are complex conjugates, this is equivalent to swapping the two saddle points.
\end{enumerate}

It is plausible \cite{RovelliDiscreteSymmetriesIn} that the two saddle-point configurations should be understood as two different values of $\sign(\det e)$. Then the $\mp$ sign in \eqref{eq:SpinfoamS_complex} reflects the fact that the real part of the on-shell first-order Holst action \eqref{eq:Holst_simple} changes sign together with $\det e$. The added dependence on $\sign(\gamma)$ in the spinfoam result is not surprising. In the spinfoam model, there is no variable corresponding to $\det e$ directly, and the difference between left-handed and right-handed orientations enters only through $\gamma$. As we've seen, the feature of two separate analytical domains (section \ref{sec:BoundaryTerm:det_e}) also survives in the spinfoam result, but with $\gamma$ as the relevant variable. Finally, we stress that \emph{all} the dependence of the amplitude on $\gamma$ is a ``quantum effect'', with no counterpart in the classical on-shell action.

\section{Discussion} \label{sec:Discussion}

In the previous sections, two main observations were made. In section \ref{sec:BoundaryTerm}, we've shown that the imaginary part of the GR action resulting from the York-Gibbons-Hawking boundary term is present also in first-order formulations. In section \ref{sec:SpinFoam}, we've shown that this imaginary part can also be recovered in a semi-classical analysis of the EPRL/FK spinfoam model, by analytically continuing the Barbero-Immirzi parameter $\gamma$ to $\pm i$ in the final result. 

The original motivation for even considering this analytical continuation stems from the recent observation \cite{FroddenBlackHoleEntropy} that the Bekenstein-Hawking entropy with the correct numerical coefficient can be obtained within loop quantum gravity by a similar procedure. We comment further on this issue in section \ref{sec:Discussion:BlackHole}. In particular, we interpret the calculation of \cite{FroddenBlackHoleEntropy}, along with our results from section \ref{sec:SpinFoam}, in the context of a ``transplanckian'' regime of LQG. For us, both calculations demonstrate that this regime correctly reproduces certain properties of semiclassical GR, provided that one sets $\gamma = \pm i$.

We stress that loop quantum gravity is well-defined only for real values of $\gamma$, so that it's currently a purely formal statement to consider a complex $\gamma$ in the quantum theory. On the other hand, an analytical continuation as the one presented could serve as an indirect definition of the theory with complex $\gamma$. While this may seem a suspicious procedure, it is tempting to follow it through. Indeed, self-dual variables corresponding to $\gamma = \pm i$ have a distinguished role already in the classical theory, leading to a polynomial (density-weight two) Hamiltonian constraint. 

\subsection{The large-spin limit as a high-energy ``transplanckian'' regime} \label{sec:Discussion:LargeSpin}

From the point of view of quantum field theory, Lagrangian parameters such as $\gamma$ are expected to run under the renormalization group (RG). We've observed that $\gamma$ has two limiting behaviors. In classical GR, the value of $\gamma$ is arbitrary, and has no effect on the action. On the other hand, in the large-spin limit of the spinfoam model, $\gamma$ must take the values $\pm i$ for the effective action to have the correct GR-like behavior, in particular the correct imaginary part. This leads us to suspect a sort of RG flow between the two regimes. To make this statement more precise, we must understand what is meant by the large-spin limit.

Before anything else, the large-spin limit of LQG is a mathematical structure. As such, it may enter physics in different contexts. In the spinfoam literature, the large-spin limit is often referred to as ``the'' semiclassical limit of LQG, with the implication that it produces the classical continuum GR that we observe at large distances. See e.g. \cite{RovelliOnTheStructure}.
In this picture, the magnitude of the spins defines a mesoscopic scale, set between the Planck scale and the scale of continuum wavelengths that one wishes to describe. At this mesoscopic scale, some set of observables, e.g. areas and angles in Regge gravity, takes the values of a chosen classical geometry, with small relative uncertainties. The continuum emerges from adding together such semiclassical discrete elements.  
A similar picture emerges in the canonical framework. There, one approximates phase-space points of classical GR by coherent states on a fixed graph, with fluctuating magnitudes of the spins. One chooses a graph with sufficiently many links and nodes so that the continuum is well-approximated, but sufficiently few so that the relative uncertainties are minimized. See chapter 11 of \cite{ThiemannModernCanonicalQuantum} and \cite{SahlmannCoherentStatesFor} for a discussion and \cite{ThiemannGCS1, ThiemannGCS2, ThiemannGCS3, ThiemannGCS4} for original literature\footnote{The usual caveats of using coherent states apply: only a chosen subalgebra of classical observables can be approximated, and the coherent states must be geared to that subalgebra. A systematic construction principle is however provided by the complexifier method \cite{ThiemannComplexifierCoherentStates}. Dynamical stability of coherent states is unknown for non-linear systems like GR. This point is especially important in the context of quantum gravity, since one would like to prove the dynamical stability of e.g. Minkowski space. In the diff-invariant context, further technical complications appear such as graph-changing operators and constraints, leading to the algebraic quantum gravity program \cite{GieselAQG1, GieselAQG2, GieselAQG3, GieselAQG4}.}.

Approaches of this kind are potentially valid, if one views graphs with large spins as a coarse-graining of finer graphs with small spins, in a spirit similar to RG flow. However, the details of such a coarse-graining procedure, in particular the link between $j\sim 1$ and $j\gg 1$, are not well-understood. Of course, if LQG is correct, there must be \emph{some} coarse-graining procedure that results in continuum GR. This procedure may or may not involve the mathematical structure of the large-spin limit.

We in this section will consider ``the'' large-spin limit in a more straightforward context - as the special subset of states in the fundamental theory that happen to have large spin labels, with no coarse-graining implied. In particular, the couplings $G$ and $\gamma$ are the same as those in the fundamental theory (which need not be the case in the coarse-graining picture). Note that this regime necessarily exists in LQG, whatever ones opinion about using large spins for coarse-graining. The ``classical GR'' that this limit reproduces is a discretized version of the GR that was quantized to obtain LQG. This is not the same as the observed GR, which is meant to emerge through coarse-graining. 

Indeed, we've already noted two differences between the two ``classical GR's''. The observed GR lives in the continuum and is not sensitive to $\gamma$. In contrast, the GR of the large-spin limit is discrete (like Regge gravity) and sensitive to $\gamma$ (unlike Regge gravity). Thus, we view the effective action at large spins as distinct from the action of the observed, continuum GR. Therefore, the requirement for the large-spin effective action to have the correct classical form \eqref{eq:ClassicalS} does not really follow from consistency with the continuum theory. Instead, we see it as a consistency check on the quantization procedure itself.

What, then, is the physical meaning of this non-coarse-grained large-spin limit? Clearly, it describes the interaction of very large ``atoms of space'', with characteristic length $L\sim \sqrt{j}$. In quantum gravity, a large geometric size may arise both in a low-energy context and in a high-energy context. Here, ``low-energy'' and ``high-energy'' should be understood as much lower or much higher than the Planck scale, which (for finite $\gamma$) is the natural scale of LQG. The low-energy context refers to particles with wavelength $\sim L$. Their characteristic action over a single wavelength is (by definition) $S\sim 1$. This leads to the inverse scaling of energy with wavelength as $E\sim 1/L$. The high-energy context, on the other hand, refers to black holes with Schwarzschild radius $\sim L$. The characteristic energy is then the black hole's mass, which scales as $E\sim L$.

We argue that the large distances associated with large spins should be viewed in a \emph{high}-energy context. Indeed, long wavelengths with low energy imply a perturbation over a background spacetime. In contrast, a large-spin state in LQG \emph{determines} the structure of spacetime, much like a heavy black hole. This is further supported by the picture of LQG intertwiners as miniature ``quantum black holes'' \cite{KrasnovBlackHolesIn}. In that picture, large spins imply a large ``black hole'', and therefore high energy in the sense of a large ``Schwarzschild radius''. Another argument is that the action of e.g. a large-spin 4-simplex scales as $S\sim j\sim L^2$, which suggests an energy $E\sim S/L\sim L$, as for a black hole.

In this way, we arrive at a picture of the large-spin limit as the regime where atoms of space with transplanckian energy interact. In particle-physics terms, the analogous situation is a scattering process of two particles with center-of-mass energy $E\gg M_{\text{planck}}$. The large-spin limit corresponds to the early stages of such a process, before the energy is dissipated into light degrees of freedom. After this dissipation, it is believed that the state should be viewed as a classical black hole, following the rules of the observed classical GR (see e.g. \cite{DvaliPhysicsOfTrans}). Before the dissipation, properly transplanckian physics takes place. There is no a-priori reason that this physics should be described by anything resembling GR. In LQG (as opposed to string theory), the fundamental theory is a quantization of GR \emph{by assumption}. As a result, through the mechanism of coherent states with large quantum numbers, fundamental transplanckian processes \emph{are} described by a (discrete) classical-GR limit. However, we stress again that this is not the same limit as the continuum GR describing classical gravity in the observed world. We stress also that the large-spin limit is ``transplanckian'' only in the sense of high energies, not in the sense of small distances. Degrees of freedom that are transplanckian in both senses at once are very likely ruled out on general grounds \cite{DvaliPhysicsOfTrans}. Note also that while we talk about transplanckian energies concentrated in single spins, this does \emph{not} imply a higher-than-Planck energy density. Indeed, since we argued that the energy scales as $E\sim L\sim \sqrt{j}$, the energy density scales as $E/L^3\sim 1/L^2\sim 1/j$.

\begin{figure}%
\centering%
\includegraphics[scale=0.75]{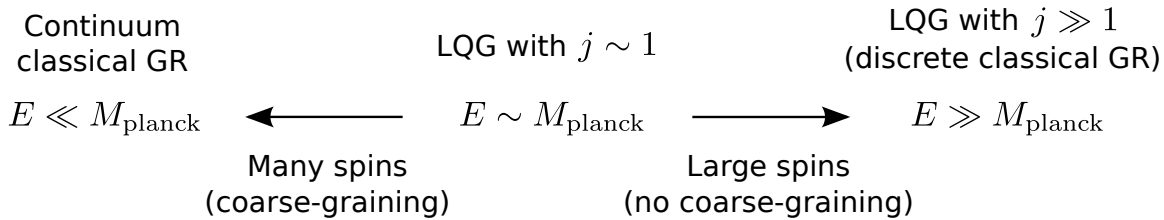} \\
\caption{Two classical-GR limits of loop quantum gravity. At high energy, discrete classical geometries are described by coherent states with large spins (no coarse graining implied). At low energy, continuum classical geometries are supposed to emerge through coarse graining.}
\label{fig:limits} 
\end{figure}

We conclude that the two classical GR's are in fact two \emph{opposite} putative limits of the quantum theory, in terms of the energies of the quanta involved. The observed continuum GR corresponds to a low-energy ``IR regime'' of subplanckian quanta (gravitons). It results from a coarse-graining procedure, which may or may not involve a large-spin description. The discrete classical GR of the large-spin limit corresponds to a high-energy ``UV regime'' of transplanckian quanta (spins and intertwiners). It results from substituting large values for the spins, with no coarse-graining implied. The situation is summarized in figure \ref{fig:limits}. It is somewhat analogous to the situation in QCD, which becomes two different free theories in the UV and in the IR (a theory of quarks and gluons in the former, and a theory of pions in the latter).

\subsection{The running of $\gamma$}

A coherent picture now emerges with respect to the running of $\gamma$. In the IR, the continuum GR is insensitive to the value of $\gamma$. In the UV, the discrete GR of the large-spin limit prefers $\gamma = \pm i$. 
This suggests that $\gamma$ ``runs'' from an arbitrary value in the IR to a UV fixed point. We stress that this ``running'' is not necessarily well-defined at the intermediate, Planck-scale energies. In the regime where individual spins are both small and relevant, it is quite possible that the dynamics is not described by any GR-like effective action. 

Interestingly, a similar picture for the flow of $\gamma$ was recently obtained in a very different framework. In \cite{BenedettiPerturbativeQuantumGravity}, Benedetti and Speziale studied the 1-loop running of $\gamma$ in effective field theory for the Holst action with a minimally coupled fermion. The calculation was done in a Euclidean framework, where the self-dual connection corresponds to $\gamma = \pm 1$ rather than $\gamma = \pm i$. It was found that $\gamma$ runs from an arbitrary initial value in the subplanckian IR to the self-dual value $\pm 1$ in the transplanckian UV. It is tempting to analytically continue this statement to the Lorentzian, saying that $\gamma$ runs from an arbitrary IR value to $\pm i$ in the UV. 

This convergence between the perturbative picture and the one suggested by large-spin amplitudes is exciting, and indeed motivated our reasoning in section \ref{sec:Discussion:LargeSpin}. Nevertheless, some obvious caveats should be mentioned. First, we stress again that we cannot directly make sense of a quantum theory with complex $\gamma$, be it in a perturbative framework or in LQG. Second, the perturbative running cannot really be trusted beyond the Planck energy. At best, it is suggestive of what should happen at transplanckian energies, with the suggestion borne out by the large-spin limit of LQG. Finally, we should note that contradictory results for the running of $\gamma$ in pure GR have been obtained in \cite{BenedettiPerturbativeQuantumGravity} in a 1-loop perturbative calculation, as well as in \cite{DaumRunningImmirziParameter, DaumEinsteinCartanGravity} in the asymptotic safety framework. However, those results refer to gauge-dependent quantities, and their significance is not clear to us.  

A last comment concerning the running is in order. As observed in section \ref{sec:BoundaryTerm:Holst}, $\gamma$ appears in the classical action as the prefactor of a combination of the Holst modification and the Nieh-Yan density. In other words, when demanding a well-defined variational principle, it is not valid to use either the Holst modification or the Nieh-Yan density individually. This point is important because one could object when that using the Nieh-Yan density as a modification for the Palatini action, the Barbero-Immirzi parameter would not run due to the topological nature of the Nieh-Yan density. This issue is also discussed from a different perspective in \cite{BenedettiPerturbativeQuantumGravity}.

\subsection{Black hole entropy at large spins with $\gamma = \pm i$} \label{sec:Discussion:BlackHole}

The black hole entropy calculation proposed in \cite{FroddenBlackHoleEntropy} has been our initial motivation for considering the substitution $\gamma = \pm i$ in the large-spin formula \eqref{eq:SpinfoamS}. In \cite{FroddenBlackHoleEntropy}, the authors propose to compute the black hole entropy within LQG in the large-spin limit by fixing the number of spin-network punctures on the horizon and setting $\gamma$ to $\pm i$. As a result, the formula for the dimension of the Chern-Simons theory describing the horizon obtains a dominant term, and the resulting entropy is given by $A/4G$. Another work in this vein is \cite{BSTI}, where the authors studied an LQG quantization of GR conformally coupled to a scalar field. There, one can count states resulting from a classical gauge-fixing of the Hamiltonian constraint. The entropy of these states comes out with a wrong dependency on the scalar field, unless one employs the reasoning of \cite{FroddenBlackHoleEntropy} and chooses $\gamma$ to reflect the self-dual case. 

The calculation in \cite{FroddenBlackHoleEntropy} is conceptually different from previous entropy calculations in LQG, see e.g. \cite{AshtekarQuantumGeometryOf}. The main difference is that in \cite{FroddenBlackHoleEntropy}, one fixes the number of punctures. From the perspective of a continuum GR limit (the low-energy ``IR limit'' of section \ref{sec:Discussion:LargeSpin}), one would expect instead to sum over all spin-network configurations with a given total area. It then follows that graphs with small spins dominate the entropy \cite{AshtekarQuantumGeometryOf}. On the other hand, in the high-energy ``UV limit'' from section \ref{sec:Discussion:LargeSpin}, it is natural to construct a ``black hole'' from a limited number of large area elements. Of course, this object does not coincide with what is usually meant by a black hole in continuum GR. However, it is the natural ``transplanckian'' analogue, given the interpretation of a continuum black hole as a large intertwiner \cite{KrasnovBlackHolesIn}. We note that the ``transplanckian'' black hole should be viewed as a temporary state: with time, it should thermalize into a continuum black hole, which in LQG will be described by many small spins. It is thus conceptually similar to the early stages of a particle collision with transplanckian center-of-mass energy.

Since the ``transplanckian'' black hole does not correspond to the continuum GR limit, there is also no need for its entropy to coincide with what is usually referred to as the Bekenstein-Hawking entropy. In fact, the entropy derived in \cite{FroddenBlackHoleEntropy,BSTI} is $A/4G$ in terms of the \emph{high}-energy Newton's constant, i.e. the bare Newton's constant $G$ from the definition of LQG. This entropy is in agreement with what one would expect from the effective (discrete) GR action in the ``transplanckian'' large-spin limit of spinfoams, which is given in terms of the same $G$. In contrast, the usual Bekenstein-Hawking entropy is given in terms of the effective low-energy $G$ of the continuum theory. There is no a priori reason why the Newton's constants in the two different classical limits should coincide. 

The entropy calculation proposed in \cite{FroddenBlackHoleEntropy} should thus be interpreted within the ``transplanckian'' regime. This is in contrast to the previous calculations such as \cite{AshtekarQuantumGeometryOf}, which aim at the continuum limit\footnote{It has been argued that the Barbero-Immirzi parameter can be fixed via black hole entropy calculations such as \cite{AshtekarQuantumGeometryOf}. This however does not seem to be the case, as newer results show \cite{EngleBlackHoleEntropyFrom, GhoshBlackHoleEntropy}. In particular, one would need to compare the result of, e.g. \cite{AshtekarQuantumGeometryOf}, with an effective continuum action derived in a proper coarse-graining procedure.}. The two main assumptions of \cite{FroddenBlackHoleEntropy}, i.e. setting $\gamma$ to $\pm i$ and considering a fixed number of spins, seem to make sense only in the ``transplanckian'' context, where they agree with the independent results from spinfoam asymptotics and from perturbative calculations. 

Together, our spinfoam calculation and the black hole state counting of \cite{FroddenBlackHoleEntropy} paint a consistent quantitative picture. There is a semiclassical limit of LQG, defined by large spins and an analytic continuation to $\gamma = \pm i$. In this limit, one can calculate an effective action from the spinfoam vertex. This action has the correct form prescribed by classical GR, including the correct relation between its real and imaginary parts. From the effective action, one can derive semiclassically the Bekenstein-Hawking entropy formula. \emph{Independently}, one can carry out a black hole state counting in the same limit, obtaining an entropy in agreement with the one derived from the action. This agreement between a statistical state counting and the Bekenstein-Hawking formula is the first of its kind within loop quantum gravity. 

\subsection{$\gamma \to \pm i$ as a result of coarse graining?}

As argued above, setting $\gamma$ to $\pm i$ for large spins has an interpretation as a ``transplanckian'' limit in the sense of large Schwarzschild radii. Now, the natural question arises whether setting $\gamma$ to $\pm i$ could also have an interpretation in terms of a (to be understood) coarse graining procedure. While we are not aware of direct arguments leading to this conclusion, we want to remark that all the above observations are also consistent with the coarse graining picture in the following (more or less vague) sense: in the coarse graining picture, the correct physical result is obtained by performing the calculation for arbitrary real $\gamma$ in the large spin limit and sending $\gamma$ to $\pm i$ after all calculations have been performed. This is especially attractive for the black hole entropy calculation, because a fixed number of (large spin) punctures would already incorporate, by the definition of coarse graining, all possible subdivisions into small spins. It is however unclear to us how sending $\gamma$ to $\pm i$ could result from a coarse graining procedure.

\section{Conclusion}

In this paper, we built on the recent observation that the second-order GR action for bounded regions has an imaginary part. We established that the same imaginary part is present also in first-order formulations of gravity. It was then shown that this imaginary part can be recovered from the large-spin asymptotics of the EPRL/FK spinfoam model, by analytically continuing the result from real Barbero-Immirzi parameter $\gamma$ to $\gamma = \pm i$. We proposed to view this limit as a ``transplanckian'' regime of LQG, in the sense of high energies but not of small distances. We argued that it's natural for the effective action in this regime to describe a discretized version of GR, governed by the high-energy Newton's constant. Further evidence for this point of view coming from a perturbative calculation and black hole entropy calculations has been discussed. While the calculations in sections \ref{sec:BoundaryTerm} and \ref{sec:SpinFoam} are robust, the arguments in the discussion section \ref{sec:Discussion} are more intuitive and should be taken with due care. Again, especially statements about the running of $\gamma$ to $\pm i$ in full Lorentzian LQG are only formal, since the quantum theory is ill-defined for non-real $\gamma$.

\section*{Acknowledgements}

NB and YN were supported by the NSF Grant PHY-1205388 and the Eberly research funds of The Pennsylvania State University. We thank Abhay Ashtekar and Marc Geiller, and Alexander Stottmeister for helpful discussions and Andreas Thurn for useful comments on a draft of this article.


\begin{thebibliography}{100}

\bibitem{NeimanOnShellActions}
Y.~Neiman, ``{On-shell actions with lightlike boundary data},'' {\tt
  arXiv:1212.2922 [hep-th]}.

\bibitem{NeimanTheImaginaryPart}
Y.~Neiman, ``{The imaginary part of the gravity action and black hole
  entropy},'' {\em Journal of High Energy Physics} {\bf 2013} (2013) 71, {\tt
  arXiv:1301.7041 [gr-qc]}.

\bibitem{YorkRoleOfConformal}
J.~W. York, ``{Role of Conformal Three-Geometry in the Dynamics of
  Gravitation},'' {\em Physical Review Letters} {\bf 28} (1972) 1082--1085.

\bibitem{GibbonsActionIntegralsAnd}
G.~W. Gibbons and S.~W. Hawking, ``{Action integrals and partition functions in
  quantum gravity},'' {\em Physical Review D} {\bf 15} (1977) 2752--2756.

\bibitem{AshtekarAsymptoticsAndHamiltoniansFour}
A.~Ashtekar, J.~Engle, and D.~Sloan, ``{Asymptotics and Hamiltonians in a
  first-order formalism},'' {\em Classical and Quantum Gravity} {\bf 25} (2008)
  095020, {\tt arXiv:0802.2527 [gr-qc]}.

\bibitem{NeimanAsymptotic}
Y.~Neiman, ``{The imaginary part of the gravitational action at asymptotic
  boundaries and horizons},'' {\tt arXiv:1305.2207 [gr-qc]}.

\bibitem{RovelliQuantumGravity}
C.~Rovelli, {\em {Quantum Gravity}}.
\newblock Cambridge University Press, Cambridge, 2004.

\bibitem{ThiemannModernCanonicalQuantum}
T.~Thiemann, {\em {Modern Canonical Quantum General Relativity}}.
\newblock Cambridge University Press, Cambridge, 2007.

\bibitem{FreidelANewSpin}
L.~Freidel and K.~Krasnov, ``{A new spin foam model for 4D gravity},'' {\em
  Classical and Quantum Gravity} {\bf 25} (2008) 125018, {\tt arXiv:0708.1595
  [gr-qc]}.

\bibitem{EngleLoopQuantumGravity}
J.~Engle, E.~R. Livine, R.~Pereira, and C.~Rovelli, ``{LQG vertex with finite
  Immirzi parameter},'' {\em Nuclear Physics B} {\bf 799} (2008) 136--149, {\tt
  arXiv:0711.0146 [gr-qc]}.

\bibitem{PerezTheSpinFoam}
A.~Perez, ``{The Spin Foam Approach to Quantum Gravity},'' {\tt arXiv:1205.2019
  [gr-qc]}.

\bibitem{HojmanParityViolationIn}
R.~Hojman, C.~Mukku, and W.~Sayed, ``{Parity violation in metric-torsion
  theories of gravitation},'' {\em Physical Review D} {\bf 22} (1980)
  1915--1921.

\bibitem{HolstBarberosHamiltonianDerived}
S.~Holst, ``{Barbero's Hamiltonian derived from a generalized Hilbert-Palatini
  action},'' {\em Physical Review D} {\bf 53} (1996) 5966--5969, {\tt
  arXiv:gr-qc/9511026}.

\bibitem{ObukhovThePalatiniPrinciple}
Y.~N. Obukhov, ``{The Palatini principle for manifold with boundary},'' {\em
  Classical and Quantum Gravity} {\bf 4} (1987) 1085--1091.

\bibitem{WielandComplexAshtekarVariables}
W.~Wieland, ``{Complex Ashtekar variables and reality conditions for Holst's
  action},'' {\em Annales Henri Poincar\'{e}} {\bf 13} (2012) 425--448, {\tt
  arXiv:1012.1738 [gr-qc]}.

\bibitem{BianchiHorizonEnergyAs}
E.~Bianchi and W.~Wieland, ``{Horizon energy as the boost boundary term in
  general relativity and loop gravity},'' {\tt arXiv:1205.5325 [gr-qc]}.

\bibitem{CorichiSurfaceTermsAsymptotics}
A.~Corichi and E.~Wilson-Ewing, ``{Surface terms, asymptotics and
  thermodynamics of the Holst action},'' {\em Classical and Quantum Gravity}
  {\bf 27} (2010) 205015, {\tt arXiv:1005.3298 [gr-qc]}.

\bibitem{MercuriFermionsInThe}
S.~Mercuri, ``{Fermions in the Ashtekar-Barbero connection formalism for
  arbitrary values of the Immirzi parameter},'' {\em Physical Review D} {\bf
  73} (2006) 084016, {\tt arXiv:gr-qc/0601013}.

\bibitem{BenedettiPerturbativeQuantumGravity}
D.~Benedetti and S.~Speziale, ``{Perturbative quantum gravity with the Immirzi
  parameter},'' {\em Journal of High Energy Physics} {\bf 2011} (2011) 1--31,
  {\tt arXiv:1104.4028 [hep-th]}.

\bibitem{FroddenBlackHoleEntropy}
E.~Frodden, M.~Geiller, K.~Noui, and A.~Perez, ``{Black Hole Entropy from
  complex Ashtekar variables},'' {\tt arXiv:1212.4060 [gr-qc]}.

\bibitem{BSTI}
N.~Bodendorfer, A.~Stottmeister, and A.~Thurn, ``{Loop quantum gravity without
  the Hamiltonian contraint},'' {\em Classical and Quantum Gravity} {\bf 30}
  (2013) 082001, {\tt arXiv:1203.6525 [gr-qc]}.

\bibitem{HawkingBlackHoleExplosions}
S.~W. Hawking, ``{Black hole explosions?},'' {\em Nature} {\bf 248} (1974)
  30--31.

\bibitem{MyersHigherDerivativeGravity}
R.~Myers, ``{Higher-derivative gravity, surface terms, and string theory},''
  {\em Physical Review D} {\bf 36} (1987) 392--396.

\bibitem{JacobsonBlackHoleEntropy}
T.~Jacobson, G.~Kang, and R.~Myers, ``{BLACK HOLE ENTROPY IN HIGHER CURVATURE
  GRAVITY},'' {\tt arXiv:gr-qc/9502009}.

\bibitem{ArnowittTheDynamicsOf}
R.~Arnowitt, S.~Deser, and C.~W. Misner, ``{The dynamics of general
  relativity},'' in {\em Gravitation: An introduction to current research}
  (L.~Witten, ed.), (New York), pp.~227--265, Wiley, 1962.
\newblock {\tt arXiv:gr-qc/0405109}.

\bibitem{DyerBoundaryTermsVariational}
E.~Dyer and K.~Hinterbichler, ``{Boundary terms, variational principles, and
  higher derivative modified gravity},'' {\em Physical Review D} {\bf 79}
  (2009) 024028, {\tt arXiv:0809.4033 [gr-qc]}.

\bibitem{HaywardGravitationalActionFor}
G.~Hayward, ``{Gravitational action for spacetimes with nonsmooth
  boundaries},'' {\em Physical Review D} {\bf 47} (1993) 3275--3280.

\bibitem{HartleBoundaryTermsIn}
J.~B. Hartle and R.~Sorkin, ``{Boundary terms in the action for the Regge
  calculus},'' {\em General Relativity and Gravitation} {\bf 13} (1981)
  541--549.

\bibitem{AshtekarNewVariablesFor}
A.~Ashtekar, ``{New Variables for Classical and Quantum Gravity},'' {\em
  Physical Review Letters} {\bf 57} (1986) 2244--2247.

\bibitem{BarberoRealAshtekarVariables}
J.~Barbero, ``{Real Ashtekar variables for Lorentzian signature space-times},''
  {\em Physical Review D} {\bf 51} (1995) 5507--5510, {\tt
  arXiv:gr-qc/9410014}.

\bibitem{RovelliTheImmirziParameter}
C.~Rovelli and T.~Thiemann, ``{The Immirzi parameter in quantum general
  relativity},'' {\em Physical Review D} {\bf 57} (1998) 1009--1014, {\tt
  arXiv:gr-qc/9705059}.

\bibitem{ImmirziQuantumGravityAnd}
G.~Immirzi, ``{Quantum gravity and Regge calculus},'' {\em Nuclear Physics B -
  Proceedings Supplements} {\bf 57} (1997) 65--72, {\tt arXiv:gr-qc/9701052}.

\bibitem{MercuriAPossibleTopological}
S.~Mercuri, ``{A possible topological interpretation of the Barbero-Immirzi
  parameter},'' {\tt arXiv:0903.2270 [gr-qc]}.

\bibitem{DateTopologicalInterpretationOf}
G.~Date, R.~Kaul, and S.~Sengupta, ``{Topological interpretation of
  Barbero-Immirzi parameter},'' {\em Physical Review D} {\bf 79} (2009) 044008,
  {\tt arXiv:0811.4496 [gr-qc]}.

\bibitem{AlexandrovTheImmirziParameter}
S.~Alexandrov, ``{The Immirzi parameter and fermions with non-minimal
  coupling},'' {\em Classical and Quantum Gravity} {\bf 25} (2008) 145012, {\tt
  arXiv:0802.1221 [gr-qc]}.

\bibitem{PeldanActionsForGravity}
P.~Peldan, ``{Actions for gravity, with generalizations: A Review},'' {\em
  Classical and Quantum Gravity} {\bf 11} (1994) 1087--1132, {\tt
  arXiv:gr-qc/9305011}.

\bibitem{RovelliDiscreteSymmetriesIn}
C.~Rovelli and E.~Wilson-Ewing, ``{Discrete symmetries in covariant loop
  quantum gravity},'' {\em Physical Review D} {\bf 86} (2012) 064002, {\tt
  arXiv:1205.0733 [gr-qc]}.

\bibitem{MikovicEffectiveActionAnd}
A.~Mikovi\'{c} and M.~Vojinovi\'{c}, ``{Effective action and semi-classical
  limit of spin-foam models},'' {\em Classical and Quantum Gravity} {\bf 28}
  (2011) 225004, {\tt arXiv:1104.1384 [gr-qc]}.

\bibitem{MikovicEffectiveActionFor}
A.~Mikovi\'{c} and M.~Vojinovi\'{c}, ``{Effective action for EPRL/FK spin foam
  models},'' {\em Journal of Physics: Conference Series} {\bf 360} (2012)
  012049, {\tt arXiv:1110.6114 [gr-qc]}.

\bibitem{EngleASpinFoam}
J.~Engle, ``{A spin-foam vertex amplitude with the correct semiclassical
  limit},'' {\em Physics Letters B} {\bf 724} (2013) 333--337, {\tt
  arXiv:1201.2187 [gr-qc]}.

\bibitem{EngleAProposedProper}
J.~Engle, ``{Proposed proper Engle-Pereira-Rovelli-Livine vertex amplitude},''
  {\em Physical Review D} {\bf 87} (2013) 084048, {\tt arXiv:1111.2865
  [gr-qc]}.

\bibitem{BianchiLorentzianSpinfoamPropagator}
E.~Bianchi and Y.~Ding, ``{Lorentzian spinfoam propagator},'' {\em Physical
  Review D} {\bf 86} (2012) 104040, {\tt arXiv:1109.6538 [gr-qc]}.

\bibitem{BarrettLorentzianSpinFoam}
J.~W. Barrett, R.~J. Dowdall, W.~J. Fairbairn, F.~Hellmann, and R.~Pereira,
  ``{Lorentzian spin foam amplitudes: graphical calculus and asymptotics},''
  {\em Classical and Quantum Gravity} {\bf 27} (2010) 165009, {\tt
  arXiv:0907.2440 [gr-qc]}.

\bibitem{NeimanParityAndReality}
Y.~Neiman, ``{Parity and reality properties of the EPRL spinfoam},'' {\em
  Classical and Quantum Gravity} {\bf 29} (2012) 065008, {\tt arXiv:1109.3946
  [gr-qc]}.

\bibitem{RovelliOnTheStructure}
C.~Rovelli, ``{On the structure of a background independent quantum theory:
  Hamilton function, transition amplitudes, classical limit and continuous
  limit},'' {\tt arXiv:1108.0832 [gr-qc]}.

\bibitem{SahlmannCoherentStatesFor}
H.~Sahlmann, T.~Thiemann, and O.~Winkler, ``{Coherent states for canonical
  quantum general relativity and the infinite tensor product extension},'' {\em
  Nuclear Physics B} {\bf 606} (2001) 401--440, {\tt arXiv:gr-qc/0102038}.

\bibitem{ThiemannGCS1}
T.~Thiemann, ``{Gauge field theory coherent states (GCS): I. General
  properties},'' {\em Classical and Quantum Gravity} {\bf 18} (2001)
  2025--2064, {\tt arXiv:hep-th/0005233}.

\bibitem{ThiemannGCS2}
T.~Thiemann and O.~Winkler, ``{Gauge field theory coherent states (GCS): II.
  Peakedness properties},'' {\em Classical and Quantum Gravity} {\bf 18} (2001)
  2561--2636, {\tt arXiv:hep-th/0005237}.

\bibitem{ThiemannGCS3}
T.~Thiemann and O.~Winkler, ``{Gauge field theory coherent states (GCS): III.
  Ehrenfest theorems},'' {\em Classical and Quantum Gravity} {\bf 18} (2001)
  4629--4681, {\tt arXiv:hep-th/0005234}.

\bibitem{ThiemannGCS4}
T.~Thiemann and O.~Winkler, ``{Gauge field theory coherent states (GCS): IV.
  Infinite tensor product and thermodynamical limit},'' {\em Classical and
  Quantum Gravity} {\bf 18} (2001) 4997--5053, {\tt arXiv:hep-th/0005235}.

\bibitem{ThiemannComplexifierCoherentStates}
T.~Thiemann, ``{Complexifier coherent states for quantum general relativity},''
  {\em Classical and Quantum Gravity} {\bf 23} (2006) 2063--2117, {\tt
  arXiv:gr-qc/0206037}.

\bibitem{GieselAQG1}
K.~Giesel and T.~Thiemann, ``{Algebraic quantum gravity (AQG): I. Conceptual
  setup},'' {\em Classical and Quantum Gravity} {\bf 24} (2007) 2465--2497,
  {\tt arXiv:gr-qc/0607099}.

\bibitem{GieselAQG2}
K.~Giesel and T.~Thiemann, ``{Algebraic quantum gravity (AQG): II.
  Semiclassical analysis},'' {\em Classical and Quantum Gravity} {\bf 24}
  (2007) 2499--2564, {\tt arXiv:gr-qc/0607100}.

\bibitem{GieselAQG3}
K.~Giesel and T.~Thiemann, ``{Algebraic quantum gravity (AQG): III.
  Semiclassical perturbation theory},'' {\em Classical and Quantum Gravity}
  {\bf 24} (2007) 2565--2588, {\tt arXiv:gr-qc/0607101}.

\bibitem{GieselAQG4}
K.~Giesel and T.~Thiemann, ``{Algebraic quantum gravity (AQG): IV. Reduced
  phase space quantization of loop quantum gravity},'' {\em Classical and
  Quantum Gravity} {\bf 27} (2010) 175009, {\tt arXiv:0711.0119 [gr-qc]}.

\bibitem{KrasnovBlackHolesIn}
K.~Krasnov and C.~Rovelli, ``{Black holes in full quantum gravity},'' {\em
  Classical and Quantum Gravity} {\bf 26} (2009) 245009, {\tt arXiv:0905.4916
  [gr-qc]}.

\bibitem{DvaliPhysicsOfTrans}
G.~Dvali, S.~Folkerts, and C.~Germani, ``{Physics of trans-Planckian
  gravity},'' {\em Physical Review D} {\bf 84} (2011) 024039, {\tt
  arXiv:1006.0984 [hep-th]}.

\bibitem{DaumRunningImmirziParameter}
J.-E. Daum and M.~Reuter, ``{Running Immirzi Parameter and Asymptotic
  Safety},'' {\tt arXiv:1111.1000 [hep-th]}.

\bibitem{DaumEinsteinCartanGravity}
J.-E. Daum and M.~Reuter, ``{Einstein-Cartan gravity, Asymptotic Safety, and
  the running Immirzi parameter},'' {\em Annals of Physics} {\bf 334} (2013)
  351--419, {\tt arXiv:1301.5135 [hep-th]}.

\bibitem{AshtekarQuantumGeometryOf}
A.~Ashtekar, J.~Baez, and K.~Krasnov, ``{Quantum Geometry of Isolated Horizons
  and Black Hole Entropy},'' {\em Advances in Theoretical and Mathematical
  Physics} {\bf 4} (2000) 1--94, {\tt arXiv:gr-qc/0005126}.

\bibitem{EngleBlackHoleEntropyFrom}
J.~Engle, K.~Noui, A.~Perez, and D.~Pranzetti, ``{Black hole entropy from an
  SU(2)-invariant formulation of Type I isolated horizons},'' {\em Physical
  Review D} {\bf 82} (2010) 044050, {\tt arXiv:1006.0634 [gr-qc]}.

\bibitem{GhoshBlackHoleEntropy}
A.~Ghosh and A.~Perez, ``{Black Hole Entropy and Isolated Horizons
  Thermodynamics},'' {\em Physical Review Letters} {\bf 107} (2011) 241301,
  {\tt arXiv:1107.1320 [gr-qc]}.



\end{thebibliography}
\end{document}